# Deep-neural-network based sinogram synthesis for sparse-view CT image reconstruction

Hoyeon Lee, Jongha Lee, Hyeongseok Kim, Byungchul Cho and Seungryong Cho, *Senior Member, IEEE*

*Abstract*— Recently, a number of approaches to low-dose computed tomography (CT) have been developed and deployed in commercialized CT scanners. Tube current reduction is perhaps the most actively explored technology with advanced image reconstruction algorithms. Sparse data sampling is another viable option to the low-dose CT, and sparse-view CT has been particularly of interest among the researchers in CT community. Since analytic image reconstruction algorithms would lead to severe image artifacts, various iterative algorithms have been developed for reconstructing images from sparsely view-sampled projection data. However, iterative algorithms take much longer computation time than the analytic algorithms, and images are usually prone to different types of image artifacts that heavily depend on the reconstruction parameters. Interpolation methods have also been utilized to fill the missing data in the sinogram of sparse-view CT thus providing synthetically full data for analytic image reconstruction. In this work, we introduce a deep-neural-network-enabled sinogram synthesis method for sparse-view CT, and show its outperformance to the existing interpolation methods and also to the iterative image reconstruction approach.

## I. INTRODUCTION

With increased use of X-ray computed tomography (CT) in clinics, potential radiation hazard has been alarmed [1, 2]. There have been developed a host of approaches toward low-dose CT imaging that include reducing or modulating the tube current, optimal selection of the tube voltage [3, 4], and sparse data sampling [5-7] to name a few. Sparse data sampling approach is in contrast with the tube current reduction since the former recruits smaller number of ray measurements with a lower noise level of the measured data than the latter. Sparse-view sampling, where the X-ray power is supposed to be turned on-and-off repeatedly, has been actively investigated as a realization of sparse data sampling although its translation to the commercialized diagnostic CT scanners has yet to come.

This project was supported in part by Korean National Research Foundation [NRF-2016M2A2A9A03913610, NRF-2016M3A9E9941837, and NRF-2017M2A2A4A05065897], and KEIT grant [10051357].

Hoyeon Lee, Jongha Lee, Hyeongseok Kim, and Seungryong Cho are with Dept. of Nuclear and Quantum Engineering, Korea Advanced Institute of Science and Engineering, Daejeon, Korea, Republic of, 34141 (e-mail: leehoy@kaist.ac.kr, jongha.lee@kaist.ac.kr, kimhs369@kaist.ac.kr, and scho@kaist.ac.kr). S. Cho is the corresponding author.

Byungchul Cho is with Dept. of Radiation Oncology, Asan Medical Center, Seoul, Korea, Republic of, 05505 (e-mail: cho.byungchul@gmail.com).

Image reconstruction from sparsely sampled data constitutes a unique, ill-posed inverse problem in CT, and the compressed-sensing-inspired algorithms have been developed to deal with this problem. Minimizing image sparsity such as image total-variation under the constraints of data fidelity and image nonnegativity has been searched for in various optimization solver frameworks [8-10]. Exploiting image sparsity in such iterative image reconstruction approaches, however, may lead to undesirable image artifacts that heavily depend on the reconstruction algorithm parameters compared to the analytically reconstructed images from fully sampled data. Additionally, the minimal amount of data that guarantees clinically acceptable image quality in various imaging tasks with varying degrees of underlying image sparsity should be carefully determined. The computation time, even though it may not constitute a critical issue with advanced acceleration techniques and parallel computing power, can still be a burden.

Direct application of analytic image reconstruction algorithm such as filtered-backprojection (FBP) to the sparse-view data would lead to images with poor quality and severe streak artifacts. Attempts have been made to synthesize the missing view data so that the full data can be fed into the analytic image reconstruction engine. An interpolation-based data synthesis in the sinogram space is a straightforward example. Various approaches have been developed for synthesizing sinogram data: linear interpolation method [11], a principal component analysis based method [12], a partial differential equation based method [13], a frequency consistency condition based method [14], intensity based directional interpolation method [15, 16], dictionary learning based method [17, 18], and some combinatorial methods [19-21]. For those interpolation approaches, image reconstruction results would highly depend on the restoring capability of the employed interpolation method. Greatly inspired by the recent progresses of machine learning techniques, we propose in this work to use a deep-neural-network for synthesizing the missing data in the sparse-view sinograms.

Machine learning has been actively used for classification tasks such as face recognition [22], tumor classification [23, 24], and image segmentation [25]. Traditional machine learning techniques train the network so that a human-defined "features" can be optimally computed in a given task such as classification and segmentation. Therefore, the performance of the machine learning heavily depends on the features that a user defines. Deep learning, on the other hand, automatically learns its own features during the network training period. The convolution layer and the activation function in the deep neural network can recruit higher dimensional features, of which a human does not have intuitive analogues, and can enhance the performance of the neural network for a variety of tasks. Thanks to such enriched capabilities of the deep learning, its applications have explosively expanded to diverse fields: for example, super-resolution image processing [26-28], visual questioning [29], generating artistic features-added photo [30], and generating new images from the random data [31]. Deep learning based approaches have also been applied to image reconstruction for low-dose CT including low tube-current CT imaging [32-34] and sparse-view CT imaging [35, 36]. While those approaches exploit the deep neural network in the image domain of CT rather than in the sinogram domain, our study focuses on restoring the missing data in the sinogram domain so that one can reconstruct images by use of the well-established reconstruction algorithms in practical uses. The advancement and wide applications of the deep learning techniques are partly stimulated by the advances in high computational power of general purpose graphic processing unit (GPGPU) and by various libraries that are publically available for individuals to apply to various fields [37-39].

In this work, we implemented a convolutional-neural-network (CNN) using the Caffe library [37] for synthesizing the missing data in the sparse-view sinogram. We would like to note that the focus of the paper is on demonstrating that a deep learning based sinogram synthesis can provide a useful solution to the low-dose CT imaging. We used real patients' CT data from The Cancer Imaging Archive (TCIA) [40] and reprojected the images to generate sinograms for training. Background of CNN, the structure of the networks used in this work, data preparation for training, and some other methods for a comparison study are described in Section 2. The comparison of the results with the other CNN approach, with the analytic interpolation methods, and also with an iterative reconstruction method will be summarized in Section 3. Discussion and conclusions will then follow.

## II. METHODS

### A. Convolutional neural network

CNN [41] is the most commonly used structure of deep neural network for imaging applications. It is composed of several layers including convolution layer, pooling layer, and fully-connected layer, and of activation functions. The convolutional layer performs convolutions to the input data with its output results forming input signals to the next layer. For each layer, weight ($W$) and bias ($b$) together with an input ($x$) are given to the layer and a convolution operation is performed as following:

$$W * x + b \qquad (1)$$

The pooling layer down-samples input data with a specific method such as maximum pooling, or average pooling. In general, the pooling layer makes shift-invariant results by maintaining specific values from the input. The shift-invariance is important in the applications such as segmentation and classification where the position of the target can be arbitrarily given in the data. Fully-connected layer refers to a layer structure in which each neuron is connected to all the neurons in the previous layer and in the next layer. Activation function is applied after fully-connected layer or convolutional layer, and it is in the form of a non-linear function such as hyperbolic tangent, sigmoid, or rectified linear unit (ReLU)[42]. With the data passing through the convolution layer, pooling layer, and activation function, the network finds features for a given task; therefore, handcrafted features are not required anymore. As training goes on, the features would evolve towards the goal with the cost function minimized. The goal in this work is to synthesize missing sinogram data, and it constitutes a kind of regression problem in which CNN-based approaches have been very successful.

### B. Structure of the proposed network

We constructed our network based on a residual U-Net. The U-Net is one of the CNN model proposed for image segmentation [43]. Residual learning is one of the techniques that can make a network converge faster and more efficiently. It trains the network to learn differences between the ground truth and the input data [26, 44]. Adding the residual learning scheme to the U-Net showed enhanced performances in removing streak artifacts in medical imaging [35, 36]. We employed the residual learning scheme in the network, and replaced pooling layers by convolutional layers to make the down-sampling trainable as well. Replacing a non-trainable layer by a trainable one has shown outperformances in other applications in the literatures [43, 45]. In our case, the measured values in the sinogram space are more important than the initial guessed values in the missing sinogram. Therefore, giving higher weights to the measured pixels or highly correlated pixels to them is more appropriate than giving higher weights to the maximum values as is often done in a max-pooling scheme. Using pooling layers is known to yield a faster output than using a stride-based convolution, since pooling does not require convolution operations. However, the restoration accuracy should not be compromised by computation time particularly in medical imaging applications; therefore, additional computation time associated with a stride-based convolution is worth taking.

By replacing pooling layers by convolutional layers, the kernel of the network will be also trained to find optimal down-sampling weights for the task. The structure of our network is shown in Fig. 1. We set the stride of the convolutional layers in association with the down-sampling to be 2, while the other layers to be 1.

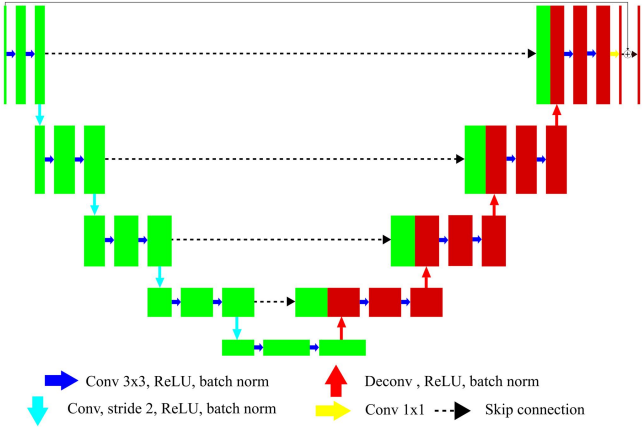

**Figure 1. Network structure for training**

Input data are prepared in patches from the sinogram as will be explained in details later and, at the bottom layer, the original input data and the output data of the last convolution layer are summed in a residual learning framework. To avoid confusion, we would like to use the dedicated terminologies throughout the paper as following: the input/output/ground truth data are used for representing the patches acquired from the input/output/ground truth sinograms, respectively. The cost function shown in the following equation compares the output data from the network with the ground truth data.

$$\frac{1}{2N}\sum_k \|x^k - y^k\|_2^2 \quad (2)$$

, where $x$ represents the network output patch, $y$ the ground truth patch in a vectorial format, $N$ the number of batches used for an iteration. The superscript $k$ refers to a patch in the training set.

### C. Training the network

Since CNN allows a supervised machine learning, we need to provide training data and ground truth data to the network. We re-projected 7 real patients' images of Lung CT[46] from The Cancer Imaging Archive (TCIA) using distance-driven projection algorithm[47] in a fan-beam CT imaging geometry in this work. The number of slice images used for training and validation was 634. The CT scan parameters are summarized in Table. I.

**Table I. Simulation conditions**

| Conditions | Value |
| --- | --- |
| Source to detector distance | 1500 mm |
| Source to axis distance | 1000 mm |
| Detector pixel spacing | 0.9 mm |
| Angular interval | 0.5 deg |
| Number of detector channels | 750 pixels |
| Number of projection data | 720 views |

We sub-sampled the original sinogram by a quarter to make them sparsely sampled one. In other words, the sparsely view-sampled sinogram with an equal angular separation between the sampled views has been prepared by selecting every forth views from the original full sinogram. Then, we applied a linear interpolation along the scan angle direction for synthesizing initial full sinogram for training the network. The same size of the input sinogram with that of the original sinogram was thus used in this work. Example images of an original full sinogram, the sparsely view-sampled sinogram, and the linearly-interpolated one are shown in Fig. 2.

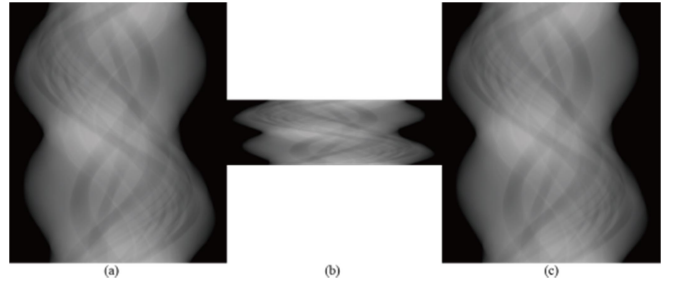

**Figure 2. (a) Ground truth sinogram, (b) Sparsely-sampled sinogram from (a), and (c) up-sampled sinogram from (b) (images from our earlier work [48])**

We would like to note again that the convolution operations have been applied to patch-based data in the CNN. Patch-based training reduces the memory requirements for input data and increases number of data used for training. From both input and ground truth sinograms, we extracted patches of the same size. We have varied the patch size and found that the patch size around 50 or bigger results in a similar network performance with bigger patch sizes requiring longer computation time due to increasing number of convolution operations. Therefore, we have extracted patches in the size of 50×50 with a stride of 10 as shown in Fig 3. A stride refers to the sampling interval in pixels between the neighboring patches. When the stride is smaller than the size of the patch, each patch has overlapping pixels with its neighboring patches. With these overlapping regions between patches, the tiling artifacts in the synthesized sinogram would be mitigated since the pixel values are averaged in the overlapped regions. The final output data of the network are still in the form of patches, which will be combined in the aforementioned way to form the synthesized sinogram.

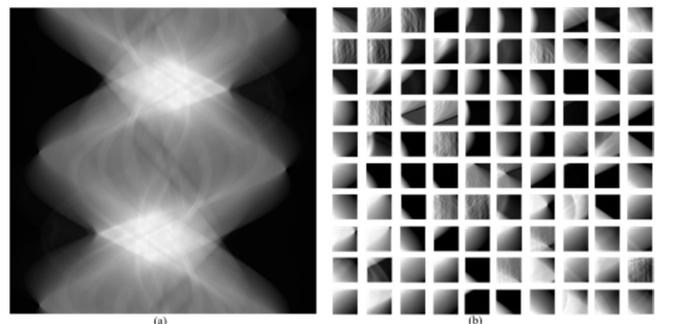

**Figure 3. (a) Up-sampled sinogram for training, and (b) patches extracted from (a)**

Our database contains 2,142,667 patches for training and 918,285 patches for validation. The training parameters of the network are summarized in Table. II. We used adaptive momentum estimation (Adam) optimizer [49] to optimize the network. It is one kind of gradient-based optimizers which has shown outperformance to the stochastic gradient descent methods. The method requires a base-learning rate, first momentum, and second momentum; we set the first and second momentum to be 0.9 and 0.999, respectively, as recommended by the original paper. The computation was done on a PC with Intel i7 2.80GHz, 16GB of random access memory, and a GPU of GTX Titan X 12GB memory.

**Table II. Training parameters**

| Conditions | Parameters |
|---|---|
| Number of data | 2,142,667 patches |
| Batch size | 67 patches |
| Learning rate | 0.0001 |
| Solver | Adam optimizer |
| Number of training epochs | 100 epochs |

### D. Other methods for comparison

For comparing the performance of the proposed method, we implemented two analytic interpolation methods and another CNN structure. A linear interpolation method and a directional interpolation method were implemented according to the literature [16]. In the linear interpolation method, a linear interpolation was performed along the angular direction to fill the missing data. The directional interpolation algorithm searches for a direction of an imaged object using the gradient of a sinogram. At a given pixel, the interpolation weights are calculated from the Eigen value and the vector of the gradient sinogram. A CNN implemented for comparison is composed of 20 successive convolutional layers with ReLU activation function, of which the structure is shown in Fig. 4. This network was used in our earlier work by the way [48].

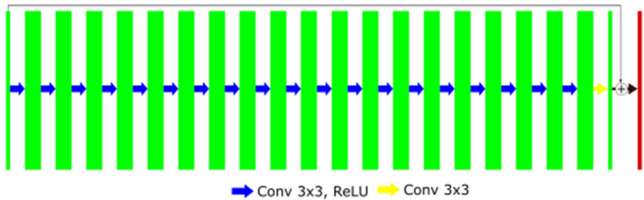

**Figure 4. 20 successive convolutional neural network**

### E. Image reconstruction

Filtered backprojection (FBP) algorithm [50] was used for image reconstruction from the ground truth sinogram and also from the synthesized sinogram. For each given imaging task, we would thus have seven FBP-reconstructed images: ground truth image, image from sparsely-sampled sinogram, images from the analytically interpolated sinograms (linear and directional interpolation), and images from the sinograms synthesized by two different deep neural networks. In addition, we have implemented an iterative image reconstruction algorithm that can directly reconstruct from the sparsely-sampled sinogram. We implemented a total variation minimization method with projection on convex sets (POCS-TV) [51].

## III. RESULTS

### A. Network training result

The training loss, or the Euclidean error of the network output of the successive convolution layers, and U-Net are plotted as a function of epochs in Fig. 5. The solid lines represent the training error, and the scatter points represent the validation error at every 20 epochs. As shown in the plots, the Euclidean loss of validation dataset has similar value to the training error for both networks. And the U-Net has resulted in a smaller error compared to the successive convolutional layers.

### B. Interpolation results

To evaluate the performance objectively, we recruited 8 patients' from the same Lung CT dataset that did not participate in training nor validation phases. The number of slices used for evaluation was 662 slices. The sinograms have been prepared in the same way according to the CT scanning geometry and separated into patches with the same size used for training, and fed into the trained networks. Two example sinograms used for evaluation are shown in Figs. 6 and 7 in their differences with the ground truth sinograms. For comparison, sinogram differences of other methods are also shown. As one can see in the figures, the synthesized sinogram by CNNs have smaller difference from the ground truth sinogram than the sinograms synthesized by other methods.

For a quantitative comparison, we computed normalized root mean-square-error (NRMSE), which is the root mean-square-error divided by the difference between maximum and minimum values of the ground truth images, peak signal-to-noise ratio (PSNR), and structural similarity (SSIM)[52]. Comparison results are summarized in Table III to Ⅴ. Although the sinogram synthesized by CNNs have similar values, the proposed network produced more accurate recovery than the successive convolutional network. This is thought to be primarily due to the fact that the proposed network synthesizes the missing data while maintaining the measured values intact.

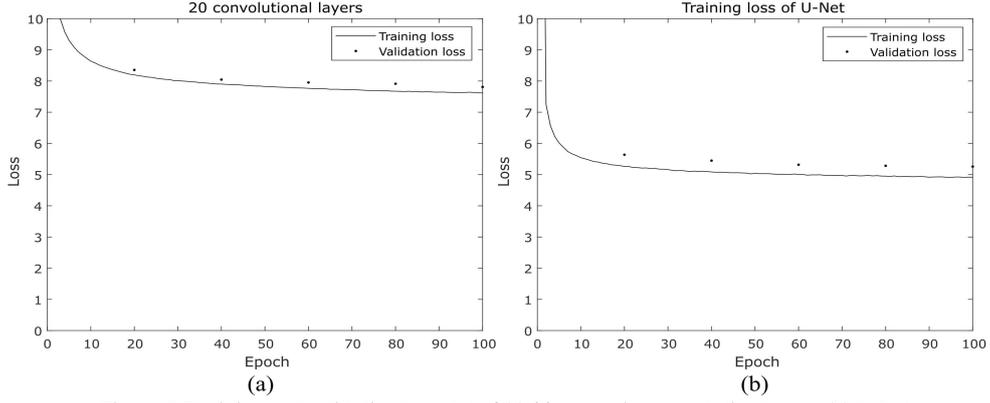

Figure 5. Training and validation loss plot of (a) 20 successive convolution layers, (b) U-Net

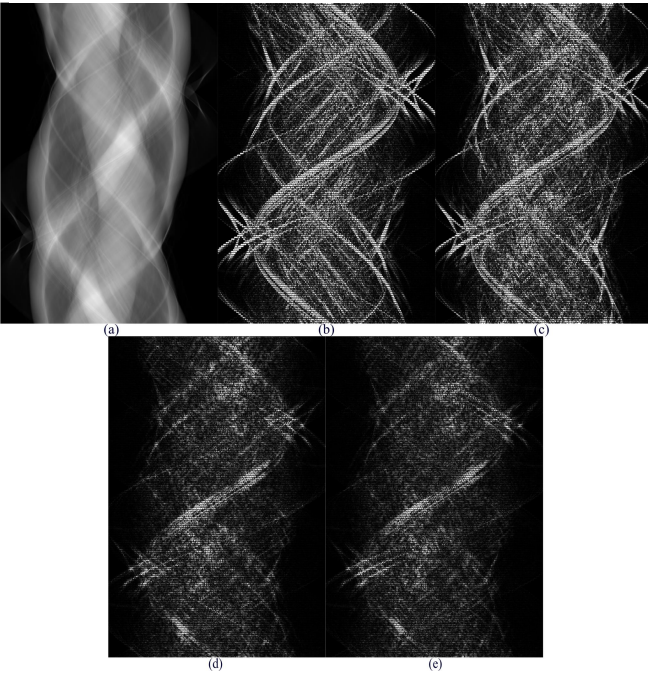

Figure 6. (a) Ground truth sinogram of patient #5, difference between ground sinogram and sinogram interpolaed using (b) linear interpolation, (c) directional interpolation, (d) 20 convolution layers, and (e) U-Net, window:[0 0.5]

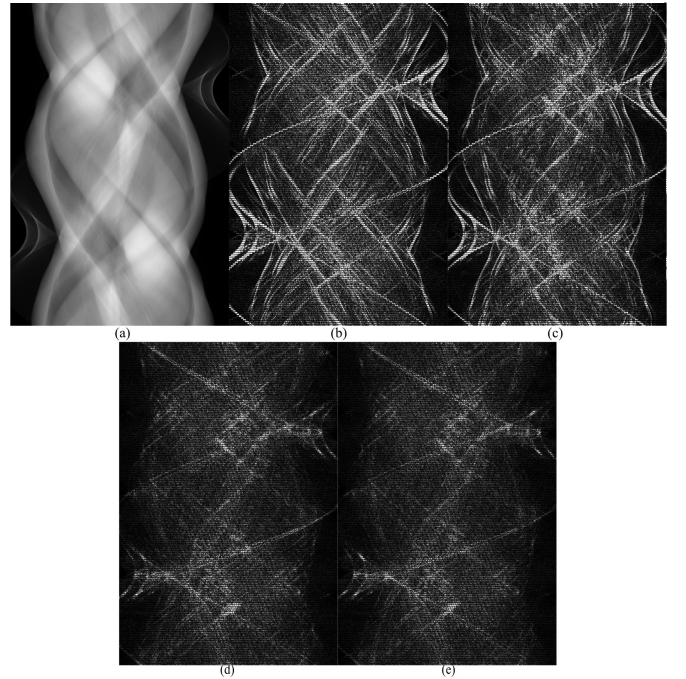

Figure 7. (a) Ground truth sinogram of patient #8, difference between ground sinogram and sinogram interpolaed using (b) linear interpolation, (c) directional interpolation, (d) 20 convolution layers, and (e) U-Net, window:[0 0.5]

**Table III. NRMSE of interpolated sinogram**

| NRMSE ($\times 10^{-3}$) | Linear interpolation | Directional interpolation | 20 Convolution | U-Net |
|---|---|---|---|---|
| Patient #1 | 2.14 | 1.73 | 0.87 | **0.79** |
| Patient #2 | 3.38 | 2.48 | 0.97 | **0.92** |
| Patient #3 | 2.19 | 1.88 | 0.95 | **0.90** |
| Patient #4 | 2.34 | 1.79 | 0.74 | **0.66** |
| Patient #5 | 2.78 | 2.32 | 1.07 | **0.96** |
| Patient #6 | 2.08 | 1.81 | 0.93 | **0.87** |
| Patient #7 | 2.56 | 2.09 | 0.92 | **0.85** |
| Patient #8 | 3.40 | 2.38 | 1.19 | **1.13** |

**Table IV. PSNR of interpolated sinogram**

| PSNR | Linear interpolation | Directional interpolation | 20 Convolution | U-Net |
|---|---|---|---|---|
| Patient #1 | 53.38 | 55.21 | 61.22 | **62.01** |
| Patient #2 | 49.32 | 52.02 | 60.18 | **60.59** |
| Patient #3 | 53.08 | 54.38 | 60.38 | **60.77** |
| Patient #4 | 52.60 | 54.95 | 62.59 | **63.59** |
| Patient #5 | 51.10 | 52.66 | 59.43 | **60.35** |
| Patient #6 | 53.52 | 54.74 | 60.49 | **61.08** |
| Patient #7 | 51.82 | 53.60 | 60.69 | **61.37** |
| Patient #8 | 49.33 | 52.44 | 58.48 | **58.94** |

**Table V. Average SSIM of interpolated sinogram**

| SSIM | Linear interpolation | Directional interpolation | 20 Convolution | U-Net |
|---|---|---|---|---|
| Patient #1 | 0.950 | 0.954 | 0.981 | **0.982** |
| Patient #2 | 0.930 | 0.941 | 0.979 | **0.981** |
| Patient #3 | 0.933 | 0.939 | 0.971 | **0.972** |
| Patient #4 | 0.947 | 0.956 | 0.983 | **0.984** |
| Patient #5 | 0.932 | 0.939 | 0.976 | **0.978** |
| Patient #6 | 0.924 | 0.935 | 0.967 | **0.970** |
| Patient #7 | 0.942 | 0.949 | 0.981 | **0.982** |
| Patient #8 | 0.865 | 0.885 | 0.925 | **0.928** |

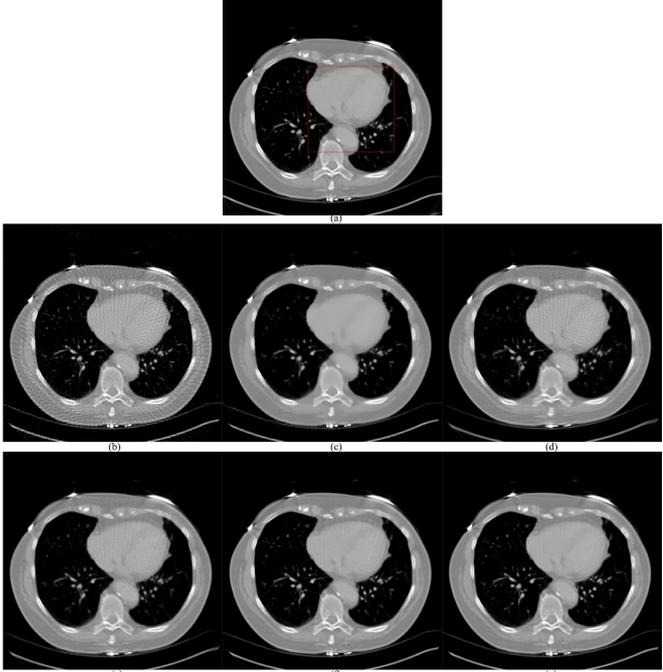

**Figure 8. Reconstructed images of (a) ground truth image, (b) sparse-view (180 views), (c) POCS-TV (180 views), (d) linear interpolation, (e) directional interpolation, (f) 20 successive convolutional layers, and (g) U-Net from patient #5, window: [-838, 593]**

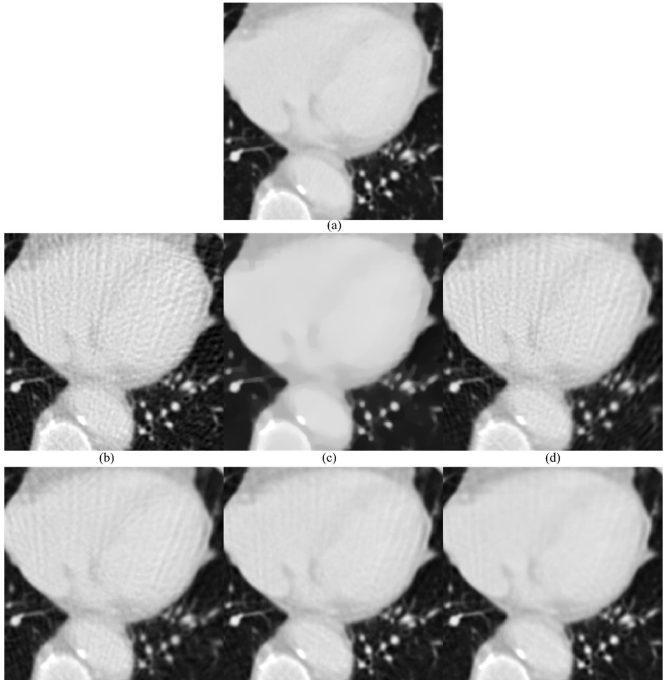

**Figure 10. Enlarged ROI of Fig. 8, window: [-1000, 300]**

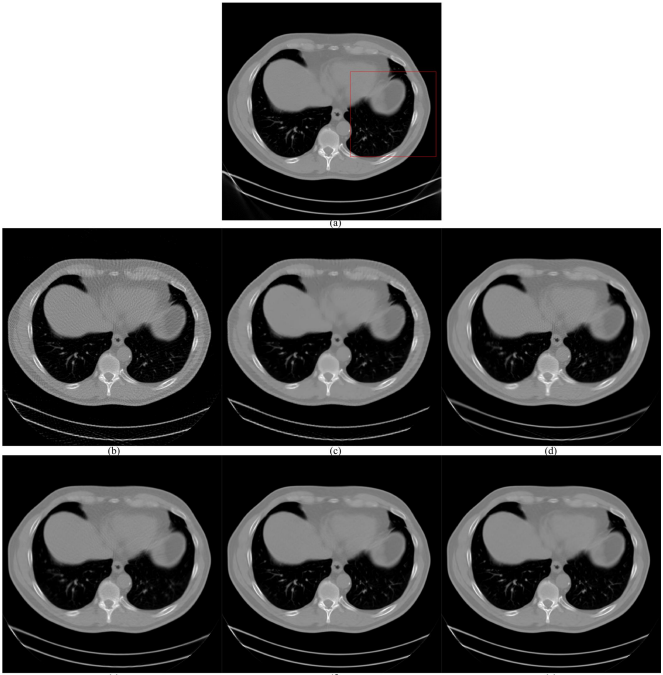

**Figure 9. Reconstructed images of (a) ground truth image, (b) sparse-view (180 views), (c) POCS-TV (180 views), (d) linear interpolation, (e) directional interpolation, (f) 20 successive convolutional layers, and (g) U-Net from patient #8, window: [-870, 760]**

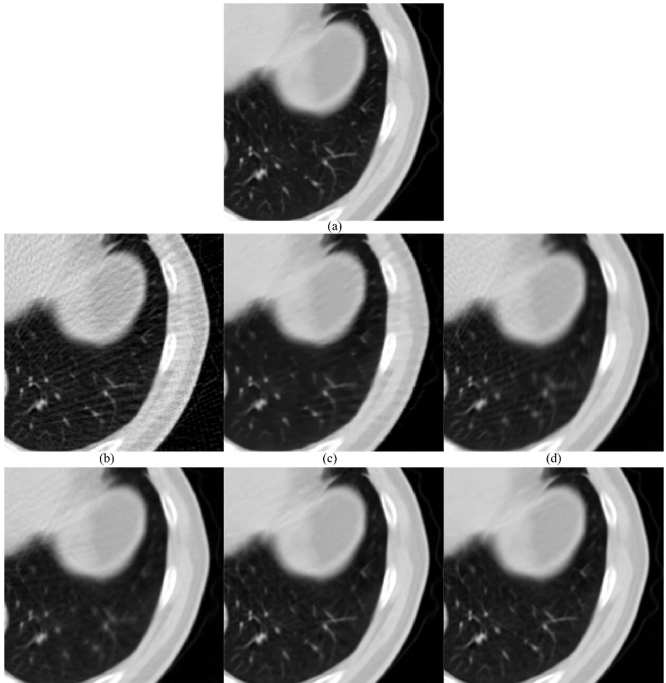

**Figure.11 Enlarged ROI of Fig. 10, window: [-1000, 285]**

Reconstructed images from the sinograms in Figs. 6 and 7 are shown in Fig. 8 and Fig. 10, respectively. For a better visual comparison, we display enlarged images of the boxed region-of-interest in Fig. 9 and Fig. 11, respectively. Quantitative comparison results, similarly to the sinogram comparison, of the reconstructed images are also summarized in Table VI to VIII.

As shown in Figs. 9 and 11, the images reconstructed by FBP algorithm directly from the sparsely sampled data suffer from severe streak artifacts. The images reconstructed by the TV minimization algorithm are subject to cartoon artifacts and they seem to miss small structures. Moderate streak artifacts remain in the images reconstructed from the synthesized sinograms by linear and directional interpolation methods. Images synthesized by CNN have smaller streak artifacts than other methods. Particularly, the reconstructed images from the sinogram synthesized by the proposed U-Net show the least streak artifacts. Results in Table VI to VIII also support the visual findings in a quantitative way; NRMSE, PSNR, and SSIM were best in the U-Net case among all the tested methods.

**Table VI. NRMSE of reconstructed images**

| NRMSE ($\times 10^{-3}$) | POCS-TV | Linear interpolation | Directional interpolation | 20 Convolution | U-Net |
|---|---|---|---|---|---|
| Patient #1 | 6.41 | 5.61 | 4.57 | 2.39 | **2.17** |
| Patient #2 | 8.04 | 10.43 | 7.52 | 2.84 | **2.73** |
| Patient #3 | 13.74 | 5.54 | 4.74 | 2.72 | **2.56** |
| Patient #4 | 6.09 | 6.18 | 4.70 | 2.05 | **1.81** |
| Patient #5 | 7.50 | 7.88 | 6.63 | 3.10 | **2.79** |
| Patient #6 | 13.60 | 5.55 | 4.65 | 2.57 | **2.39** |
| Patient #7 | 7.56 | 7.17 | 5.91 | 2.77 | **2.49** |
| Patient #8 | 12.58 | 13.43 | 9.24 | 4.46 | **4.18** |

**Table VII. PSNR of reconstructed images**

| | POCS-TV | Linear interpolation | Directional interpolation | 20 Convolution | U-Net |
|---|---|---|---|---|---|
| Patient #1 | 41.20 | 42.36 | 44.14 | 49.77 | **50.62** |
| Patient #2 | 38.62 | 36.36 | 39.20 | 47.66 | **47.99** |
| Patient #3 | 34.00 | 41.88 | 43.24 | 48.06 | **48.60** |
| Patient #4 | 41.71 | 41.58 | 43.95 | 51.18 | **52.23** |
| Patient #5 | 39.89 | 39.47 | 40.96 | 47.56 | **48.49** |
| Patient #6 | 34.70 | 42.49 | 44.03 | 49.18 | **49.79** |
| Patient #7 | 37.22 | 39.67 | 41.36 | 47.94 | **48.87** |
| Patient #8 | 33.26 | 32.69 | 35.94 | 42.26 | **42.83** |

**Table VIII. Average SSIM of reconstructed images**

| | POCS-TV | Linear interpolation | Directional interpolation | 20 Convolution | U-Net |
|---|---|---|---|---|---|
| Patient #1 | 0.960 | 0.970 | 0.974 | 0.989 | **0.991** |
| Patient #2 | 0.950 | 0.942 | 0.964 | 0.991 | **0.992** |
| Patient #3 | 0.949 | 0.969 | 0.975 | 0.988 | **0.989** |
| Patient #4 | 0.962 | 0.964 | 0.974 | 0.992 | **0.993** |
| Patient #5 | 0.956 | 0.958 | 0.966 | 0.987 | **0.990** |
| Patient #6 | 0.932 | 0.971 | 0.977 | 0.990 | **0.991** |
| Patient #7 | 0.967 | 0.969 | 0.975 | 0.990 | **0.992** |
| Patient #8 | 0.936 | 0.936 | 0.958 | 0.982 | **0.984** |

## IV. DISCUSSION

Our study reveals that the CNN-based interpolation or synthesis of the sparsely-sampled sinogram can effectively make up the missing data and can produce reconstructed images of comparable quality to the ones reconstructed from the fully-sampled sinogram. Although the POCS-TV reconstruction results from the sparsely-sampled data are rather poor in this study, we would like to note that such an iterative algorithm strongly depends on the optimization cost function and reconstruction parameters. Therefore, we cannot exclude a chance that a fine-tuned iterative algorithm can produce a reasonably acceptable image quality in a given imaging task. However, it is a common understanding that such compressed-sensing-inspired algorithms are in general subject to cartoon image artifacts and that they may miss small structures in the reconstructed images from the sparsely sampled data.

The training took about 5 days for the successive convolutional layers, and 12 days for the U-Net in our computing environment. However, the sinogram synthesis after the networks complete the training took less than 10 seconds and 50 seconds for the successive convolutional layers and the U-Net, respectively. Since we used relatively a small stride while making the training dataset, the data redundancy in the dataset is relatively high. While it helps increasing the number of training data, it also increases the training time. We will further investigate on reducing redundancies of the dataset as an attempt to increase the training speed in the future without compromised performance of the network. Based on our preliminary study, we will continue investigating the utility of the deep network in clinical environments that include cone-beam CT and helical multiple fan-beam CT. Additionally, an irregular angular sampling in the sparse-view data acquisition as well as handling missing detector channel problem would be our future study.

## V. CONCLUSION

In this study, we developed a U-Net structure for interpolating sparsely-sampled singoram to reconstruct CT images by an FBP algorithm. We trained the network with the re-projected data from the real patients' CT images. We compared the performance of the proposed method to the linearly interpolated sinogram, the directionally interpolated sinogram, and the interpolated sinograms using the other CNN. Reconstructed images have also been compared likewise, and the reconstructed image by the proposed method was also compared to the image reconstructed by a TV-minimization algorithm directly from the sparsely-sampled data. The proposed network produced promising results and is believed to play an important role as an option to the low-dose CT imaging.